\documentclass[aps,prl,amsmath,amssymb,superscriptaddress,lengthcheck]{revtex4-1}

\newcommand{\ncco}{$\textrm{Nd}_{2-x}\textrm{Ce}_x\textrm{Cu}\textrm{O}_{4}$}
\newcommand{\lcco}{$\textrm{La}_{2-x}\textrm{Ce}_x\textrm{Cu}\textrm{O}_{4}$}

\newcommand{\ybco}{$\textrm{Y}\textrm{Ba}_2\textrm{Cu}_3\textrm{O}_{6+\delta}$}
\newcommand{\lbsco}{$\textrm{La}_{2-\delta}\textrm{(Ba,Sr)}_\delta \textrm{Cu}\textrm{O}_{4}$}
\newcommand{\lsco}{$\textrm{La}_{2-\delta}\textrm{Sr}_\delta \textrm{Cu}\textrm{O}_{4}$}
\newcommand{\hg}{$\textrm{Hg}\textrm{Ba}_2\textrm{Cu}\textrm{O}_{4+\delta}$}
\newcommand{\qco}{$Q_{CO}(x)$}
\usepackage{graphicx}
\usepackage{dcolumn}
\usepackage{bm}
\usepackage{relsize}
\begin{document}

\title{Doping dependent charge order correlations in electron-doped cuprates}

\author{E.\,H.\,da Silva Neto}
\email[]{ehda@physics.ubc.ca}
\affiliation{Department of Physics \& Astronomy, University of British Columbia, Vancouver British Columbia, V6T 1Z1, Canada}
\affiliation{Quantum Matter Institute, University of British Columbia, Vancouver, British Columbia V6T 1Z4, Canada}
\affiliation{Max Planck Institute for Solid State Research, Heisenbergstrasse 1, D-70569 Stuttgart, Germany}
\affiliation{Quantum Materials Program, Canadian Institute for Advanced Research, Toronto, Ontario M5G 1Z8, Canada.}

\author{B.\,Yu}
\affiliation{School of Physics and Astronomy, University of Minnesota, Minneapolis, Minnesota 55455, USA}

\author{M.\,Minola}
\affiliation{Max Planck Institute for Solid State Research, Heisenbergstrasse 1, D-70569 Stuttgart, Germany}

\author{R.\,Sutarto}
\affiliation{Canadian Light Source, Saskatoon, Saskatchewan S7N 2V3, Canada}

\author{E.\,Schierle}
\affiliation{\mbox{Helmholtz-Zentrum Berlin für Materialien und Energie, Albert-Einstein-Strasse 15, D-12489 Berlin, Germany.}}

\author{\mbox{F.\,Boschini}}
\affiliation{Department of Physics \& Astronomy, University of British Columbia, Vancouver British Columbia, V6T 1Z1, Canada}
\affiliation{Quantum Matter Institute, University of British Columbia, Vancouver, British Columbia V6T 1Z4, Canada}

\author{M.\,Zonno}
\affiliation{Department of Physics \& Astronomy, University of British Columbia, Vancouver British Columbia, V6T 1Z1, Canada}
\affiliation{Quantum Matter Institute, University of British Columbia, Vancouver, British Columbia V6T 1Z4, Canada}

\author{M.\,Bluschke}
\affiliation{Max Planck Institute for Solid State Research, Heisenbergstrasse 1, D-70569 Stuttgart, Germany}
\affiliation{\mbox{Helmholtz-Zentrum Berlin für Materialien und Energie, Albert-Einstein-Strasse 15, D-12489 Berlin, Germany.}}

\author{J.\,Higgins}
\affiliation{\mbox{Center for Nanophysics and Advanced Materials, University of Maryland, College Park, MD 20742, USA.}}

\author{Y.\,Li}
\affiliation{School of Physics and Astronomy, University of Minnesota, Minneapolis, Minnesota 55455, USA}

\author{G.\,Yu}
\affiliation{School of Physics and Astronomy, University of Minnesota, Minneapolis, Minnesota 55455, USA}

\author{E.\,Weschke}
\affiliation{\mbox{Helmholtz-Zentrum Berlin für Materialien und Energie, Albert-Einstein-Strasse 15, D-12489 Berlin, Germany.}}

\author{F.\,He}
\affiliation{Canadian Light Source, Saskatoon, Saskatchewan S7N 2V3, Canada}

\author{M.\,Le\,Tacon}
\affiliation{Max Planck Institute for Solid State Research, Heisenbergstrasse 1, D-70569 Stuttgart, Germany}
\affiliation{Institut f\"ur Festkörperphysik, Karlsruher Institut f\"ur Technologie, 76201 Karlsruhe, Germany}

\author{R.\,L.\,Greene}
\affiliation{\mbox{Center for Nanophysics and Advanced Materials, University of Maryland, College Park, MD 20742, USA.}}

\author{M.\,Greven}
\affiliation{School of Physics and Astronomy, University of Minnesota, Minneapolis, Minnesota 55455, USA}

\author{G.\,A.\,Sawatzky}
\affiliation{Department of Physics \& Astronomy, University of British Columbia, Vancouver British Columbia, V6T 1Z1, Canada}
\affiliation{Quantum Matter Institute, University of British Columbia, Vancouver, British Columbia V6T 1Z4, Canada}

\author{B.\,Keimer}
\affiliation{Max Planck Institute for Solid State Research, Heisenbergstrasse 1, D-70569 Stuttgart, Germany}

\author{A.\,Damascelli}
\email[]{damascelli@physics.ubc.ca}
\affiliation{Department of Physics \& Astronomy, University of British Columbia, Vancouver British Columbia, V6T 1Z1, Canada}
\affiliation{Quantum Matter Institute, University of British Columbia, Vancouver, British Columbia V6T 1Z4, Canada}


\begin{abstract}
Understanding the interplay between charge order (CO) and other phenomena (e.g. pseudogap, antiferromagnetism, and superconductivity) is one of the central questions in the cuprate high-temperature superconductors. The discovery that similar forms of CO exist in both hole- and electron-doped cuprates opened a path to determine what subset of the CO phenomenology is universal to all the cuprates. Here, we use resonant x-ray scattering to measure the charge order correlations in electron-doped cuprates (\lcco~and \ncco) and their relationship to antiferromagnetism, pseudogap, and superconductivity. Detailed measurements of \ncco~show that CO is present in the $x = 0.059$ to $0.166$ range, and that its doping dependent wavevector is consistent with the separation between straight segments of the Fermi surface. The CO onset temperature is highest between $x = 0.106$ and $0.166$, but decreases at lower doping levels, indicating that it is not tied to the appearance of antiferromagnetic correlations or the pseudogap. Near optimal doping, where the CO wavevector is also consistent with a previously observed phonon anomaly, measurements of the CO below and above the superconducting transition temperature, or in a magnetic field, show that the CO is insensitive to superconductivity. Overall these findings indicate that, while verified in the electron-doped cuprates, material-dependent details determine whether the CO correlations acquire sufficient strength to compete for the ground state of the cuprates.

\end{abstract}

\maketitle

\begin{figure*}
\includegraphics[width=160mm]{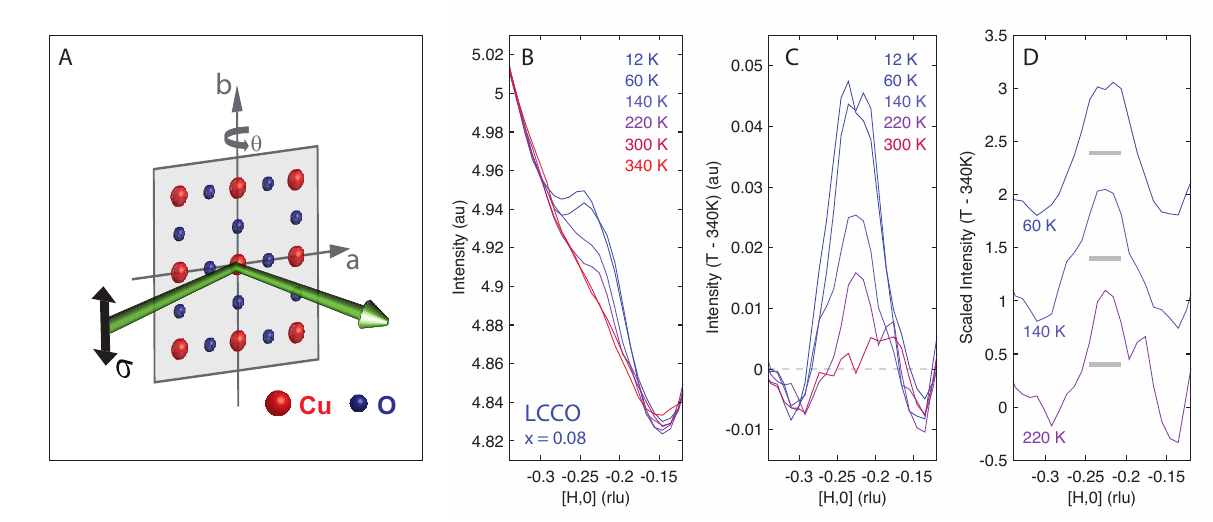}
\caption{\label{fig:1} \textbf{Charge order in LCCO.} (\textbf{A}) Scattering geometry along the Cu-O bond direction (see text for details). (\textbf{B}) RXS $\theta$-scans for LCCO ($x=0.08$, $T_c \sim 20$\,K) at various temperatures. (\textbf{C}) CO peaks at different temperatures obtained upon subtracting the $340$\,K data from those at lower temperatures. (\textbf{D}) $60$\,K, $140$\,K, $220$\,K data from (C) with their maxima normalized to unit. The curves were offset for clarity and the width of gray bars represents the half-width-at-half-maximum of the $60$\,K data.}
\end{figure*}

\section{Introduction}

In the copper oxide materials the doping of either hole- or electron-carriers into the parent Mott insulator suppresses antiferromagnetism and allows the appearance of superconductivity \cite{armitage_progress_2010}. Several studies have shown that the cuprates also feature a tendency toward a periodic self-organization of their charge degrees of freedom, known as charge order (CO) \cite{tranquada_evidence_1995, Abbamonte_2005, Wu_2011, Ghiringhelli_CDW_2012, Chang_CDW_2012, Achkar_2012, Comin_CDW_2014, daSilvaNeto_CDW_2014, Hashimoto_2014, Tabis_2014, daSilvaNeto_Science_2015, Yazdani_ARCMP, Comin_ARCMP}. Though CO is now accepted as a ubiquitous phenomenon in the cuprates, the situation regarding its interrelationship with antiferromagnetism (AFM), superconductivity, and the pseudogap is less clear. For example, in the La-based hole cuprates, the CO periodicity increases as a function of doping and is closely related to the antiferromagnetic incommensurability near the hole doping level $1/8$ \cite{tranquada_evidence_1995, Yamada_1998}, but this connection is absent in other cuprates \cite{Comin_CDW_2014, daSilvaNeto_CDW_2014, Tabis_2014, Santi_2014, Huecker_2014, Chan_2016}. Regarding the pseudogap, though x-ray and STM measurements show that the CO onsets below, or at the pseudogap temperature (T*) \cite{parker_fluctuating_2010, Comin_CDW_2014, Santi_2014, Huecker_2014} in hole-doped cuprates, opposite behavior is observed in electron-doped cuprates \cite{daSilvaNeto_Science_2015}. Finally, several studies point toward a competition between CO and superconductivity \cite{tranquada_evidence_1995, Abbamonte_2005, Ghiringhelli_CDW_2012, Chang_CDW_2012, daSilvaNeto_CDW_2014, Moodenbaugh_1988, Buchner_1994, Croft_PRB_2014, Christensen_2014}, though this has not yet been experimentally verified in electron-doped cuprates due to the lack of measurements across the superconducting transition temperature ($T_c$), as a function of doping, or in the presence of a magnetic field.

Despite the wealth of experimental studies of CO in the hole-doped cuprates, a comprehensive study of the electron-doped counterparts has not yet been reported.  If CO is found to be a universal property of the electron- doped cuprates as well, detailed knowledge of its behavior as a function of doping, temperature and magnetic field can be used not only to determine which emergent phenomena (i.e. superconductivity, pseudogap, and antiferromagnetism) are truly intrinsically connected to CO formation, but also to provide clues about its microscopic origin. For example, due to the robustness of antiferromagnetic correlations in electron-doped cuprates relative to the hole-doped materials, the study of the CO in the former might serve as a test of several theoretical models in which charge order is magnetically driven \cite{Davis_DHLee_2013, Efetov_nat_phys_2013, Sachdev_PRL_2013, Wang_Chubukov_PRB_2014}. Indeed, it has been shown that for \ncco~(NCCO) near optimal doping, the CO wave vector ($Q_{CO}$) connects the areas of the Brillouin zone near where the AFM zone boundary intercepts the underlying Fermi surface, similar to observations in Bi-based hole-doped cuprates \cite{Comin_CDW_2014, daSilvaNeto_CDW_2014, Fujita_Science_2014}. Further supporting this connection between CO and AFM, it has been noted that the temperature scales for the two phenomena approximately coincide for NCCO near optimal doping. However, these relationships between CO and AFM in momentum space and in temperature have not yet been studied as a function of doping. Such comprehensive doping-temperature study of the CO should not only clarify its connection to AFM, but also to the pseudogap phenomena, which in electron-doped cuprates is thought to derive from the AFM \cite{armitage_progress_2010}.

\section{Results}

Before investigating the universality of the CO phenomenology, we first demonstrate the presence of CO correlations in a second member of the family of electron-doped family, \lcco~(LCCO). To reveal the CO correlations in this material, we performed resonant x-ray scattering (RXS) measurements at the \mbox{Cu-$L_3$} edge. The in-plane components of momentum transfer along the \mbox{$a$-axis} were accessed by rotating the sample about the \mbox{$b$-axis} ($\theta$-scan), while maintaining the scattering geometry (angles of incoming and outgoing photons) fixed, as depicted in Fig.\,\ref{fig:1}A. All momentum transfer components $\vec{Q}=(H, K, L)$ are reported in reciprocal lattice units~(rlu). Figures\,\ref{fig:1}B-D show our RXS measurements of LCCO, performed at various temperatures. It is clear from the \mbox{$\theta$-scans} in Fig.\,\ref{fig:1}B, that a peak near $H = -0.22$ is present at $12$\,K, weakens as the temperature is raised, and disappears above $220$\,K. The suppression of this peak either with temperature, or by tuning the photon energy away from resonance \cite{SM}, demonstrates the presence of CO in LCCO and validates the presence of charge order in electron-doped cuprates. Unlike previous work for NCCO \cite{daSilvaNeto_Science_2015}, the data in LCCO show a featureless high-temperature \mbox{$\theta$-scan}, which provides a good measure of the temperature-independent background. Subtraction of the $340$\,K data reveals a CO peak with similar width ($\sim$\,$25$\,\AA~correlation length in real space) and intensity as for NCCO (see Fig.\,\ref{fig:1}C). Furthermore it also reveals that the peak width is remarkably temperature-independent, as can be seen more clearly in Fig.\,\ref{fig:1}D which shows the curves in Fig.\,\ref{fig:1}C normalized by their maxima. As we will discuss below, the temperature-independence of the CO correlation length also appears to be present in NCCO and provides a key difference between the phenomenologies of \ybco~(YBCO), as well as \lbsco, and electron-doped cuprates. 

\begin{figure*}
\includegraphics[width=182mm]{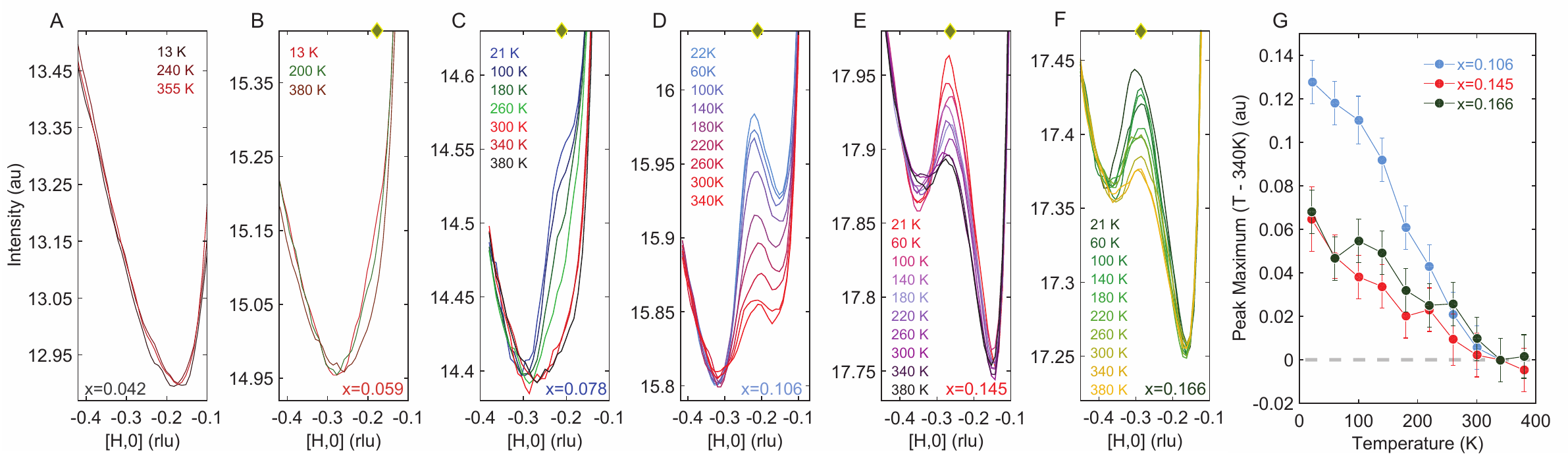}
\caption{\label{fig:2} \textbf{Temperature and doping dependence of CO in NCCO.} (\textbf{A-F}) Temperature dependence of $\theta$-scans for six doping levels of NCCO. Yellow diamonds in (A-F) show the $H$ location of the low temperature peak maxima \cite{SM}. (\textbf{G}) Temperature dependence of the CO peak maximum after subtraction of the $340$\,K peak maximum \cite{SM}, extracted from the the data in (D-F). The vertical scales in (A-F) are proportional to the detector reading normalized to the incoming photon flux \cite{SM}. Note that the intensity difference in the vertical scale of (G) is plotted in the same units as in (D-F). The error bars in (G) represent the systematic errors associated with the experiment \cite{SM}.}
\end{figure*}

\begin{figure}
\includegraphics[width=80mm]{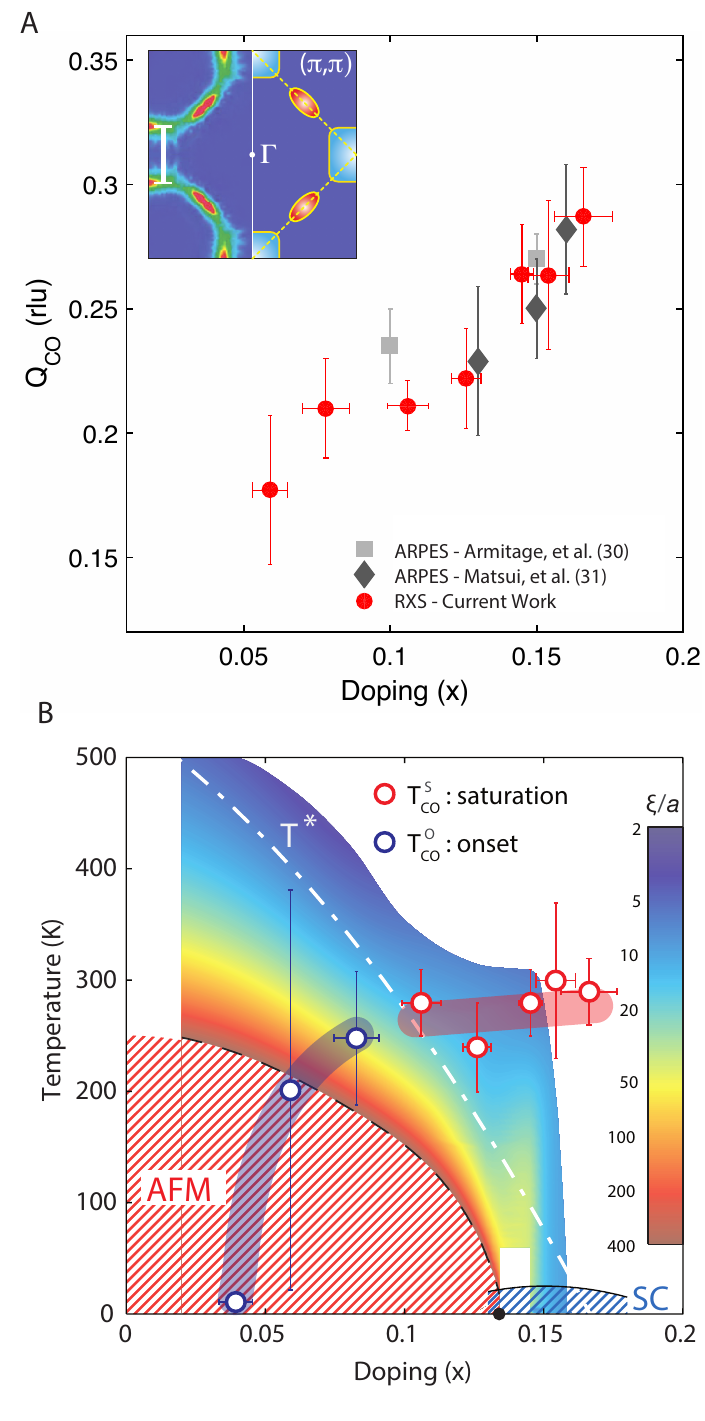}
\caption{\label{fig:3} \textbf{Phase diagram of CO in NCCO.} (\textbf{A}) Doping dependence of \qco~compared with the separation between the segments of the Fermi surface near $(\pi, 0)$ as determined from ARPES (white arrow in the inset). The inset shows a representative ARPES Fermi surface NCCO for $x=0.15$ (left) and a schematic of the AFM-folded Fermi surface (right), with electron (blue) and hole (red) pockets. (\textbf{B}) Phase diagram of NCCO adapted from \cite{Motoyama_Spin_2007}, including the antiferromagnetism (AFM) and superconductivity (SC) region, the pseudogap temperature ($T^*$), and the instantaneous AFM correlation length ($\xi$) (normalized to the tetragonal lattice constant $a$) determined via inelastic neutron scattering. Superimposed red and blue circles represent $T_{CO}^O$, and $T_{CO}^S$, respectively. Thick semi-transparent blue and red lines are guides to the eye. In (A-B) the horizontal error bars represent the uncertainty in the experimental determination of doping level \cite{SM}. The vertical error bars in (B) indicate the uncertainty in locating the temperature where $T_{CO}^O$, and $T_{CO}^S$ deviate from their respective high-temperature behaviors \cite{SM}.}
\end{figure}

Figures\,\ref{fig:2}A-F show $\theta$-scans at a number of temperatures for six different doping levels of NCCO. It is clear from these measurements that the CO is absent for $x = 0.042$ at low temperatures. Measurements above the N\'eel temperature (see Fig.\,\ref{fig:2}A) indicate that doping, rather than competition for the ground state, is responsible for the suppression of the CO at this doping range. Figures\,\ref{fig:2}C-E also explicitly show that \qco~increases as a function of~$x$. In Fig.\,\ref{fig:3}A we summarize \qco~and, from comparison to previous ARPES reports \cite{Armitage_doping_2002, Matsui_evolution_2007}, show that it connects the parallel segments near $(\pi,0)$. The value of \qco~for $x=0.14$ is also consistent with inelastic hard x-ray scattering measurements that detect an anomalous softening of the bond stretching mode at $H = 0.20 \pm 0.03$ \cite{dAstuto_2002}. Here we observe that due to the Fermi surface topology, we cannot rule out that $Q_{CO}(x)$ connects the inter-hot-spot distance \cite{SM}, although, as we show below (Fig.\,\ref{fig:3}B), the combined doping-temperature dependence of the CO seems uncorrelated to $T^*$ \cite{daSilvaNeto_Science_2015}. While this observation, together with the absence of gaps near $(\pi,0)$ above $T_c$, suggest that Fermi surface nesting is not responsible for the CO formation, it is possible that the CO never becomes sufficiently long-range, or intense enough, to modify the Fermi surface. Nevertheless, note that a similar behavior of the CO peak is observed in YBCO and Bi-cuprates, where \qco~also follows the Fermi surface \cite{Comin_CDW_2014, daSilvaNeto_CDW_2014, Santi_2014}. Therefore, our finding for \qco~adds to the list of similarities between NCCO and hole-doped COs -- in addition to the similar wavevector, coherence length and RXS intensity (the last two in Bi-cuprates) -- and further supports a common origin for their existence.

The possible connection between \qco~and the inter-hot-spot distance suggests that perhaps antiferromagnetic fluctuations are intrinsically connected to the appearance of CO instabilities. To explore this idea we turn to the temperature dependence of the CO for NCCO and compare it to previous inelastic neutron scattering (INS) studies that probe the instantaneous antiferromagnetic fluctuations \cite{Motoyama_Spin_2007}. To ensure a direct experimental comparison, we used crystals that were either obtained from larger pieces used in the previous INS study \cite{Motoyama_Spin_2007} or synthesized by the same method \cite{SM}. For \mbox{$x=0.059$}, a weak CO peak is barely detectable, as shown in Fig.\,\ref{fig:2}B, and completely disappears above room-temperature. The relative weakness of the CO peak for this sample precludes a precise measurement of its onset temperature. Upon further doping the $x=0.078$ sample shows a clearer CO peak which, at $T_{CO}^{O} = 260 \pm 60$\,K, also completely disappears. However, this behavior qualitatively changes for \mbox{$x = 0.106$}, where the CO peak apparently saturates at a finite value, $T_{CO}^{S}$, above room-temperature, as shown in Figs.\,\ref{fig:2}D-F. The reason for this behavior is not clear, and we cannot distinguish whether it is truly a saturation or a change in the rate at which the CO is suppressed with temperature -- the latter would require measurements at temperatures higher than what is currently technically possible. This temperature dependence is summarized in Fig.\,\ref{fig:2}G, where the peak maximum (after subtracting the $340$\,K data) is shown for the data in Fig.\,\ref{fig:2}D-F. Unfortunately, the high-temperature behavior, discussed above, precludes the determination of the true background and renders any determination of the CO intensity versus doping unreliable \cite{SM}. Nevertheless, it is obvious that above $x=0.042$ the CO temperature scale increases up to $x=0.106$, and remains high (above $300$\,K) with further electron-doping. The behavior of the characteristic temperatures shows a trend opposite to the antiferromagnetic correlations, as summarized in Fig.\,\ref{fig:3}B, and suggests that the two phenomena are not intrinsically related.

For a typical second-order phase transition, the correlation length and susceptibility increase upon cooling in the disordered phase. Such behavior of the CO is seen in YBCO above $T_c$. However, as Fig.\,\ref{fig:1} shows, the CO peak width in LCCO is temperature independent, and the correlation length never increases above $\sim$\,$25$\,\AA. At first sight, such a clear assertion cannot be made about the correlation length of the CO in NCCO since the presence of a peak at all measured temperatures precludes the identification of the true background. Nevertheless, the data in Fig.\,\ref{fig:2}D-F show that the peak develops on top of a concave fluorescence background, displaying a distinct local minimum for $H$ less than $Q_{CO}$. Under these conditions, this minimum should move away from $Q_{CO}$ if the width increased with temperature -- a behavior which is clearly not present in the data \cite{SM}. Therefore, though a precise measure of the peak width as a function of temperature cannot be obtained, we can conclude that, as in LCCO, and contrary to YBCO, the correlation length in NCCO is approximately independent of temperature. This behavior, as well as the short correlation length, resemble the observations for hole-doped Bi-cuprates \cite{Comin_CDW_2014, daSilvaNeto_CDW_2014} and \hg~(Hg1201) \cite{Tabis_2014}.

\begin{figure}
\includegraphics[width=80mm]{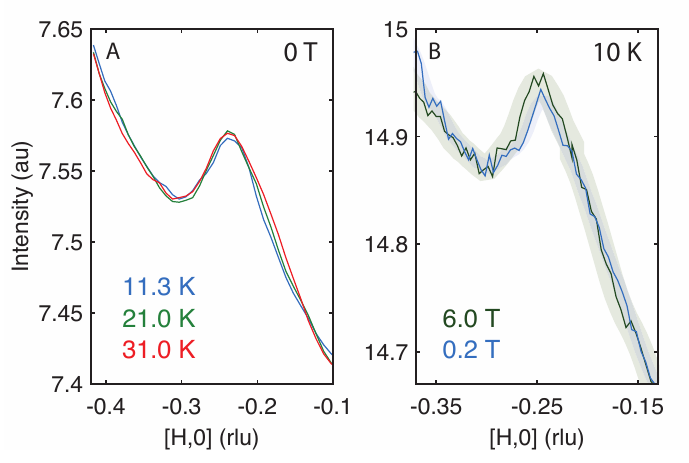}
\caption{\label{fig:4} \textbf{Charge order versus superconductivity.} Measurements of an $x=0.14$ sample ($T_c \sim 22$\,K) (\textbf{A}) at $0$\,T as a function of temperature above and below $T_c$, and (\textbf{B}) at $10$\,K for $0.2$ and $6.0$\,T. The halos around the curves in (B) represent the experimental uncertainty from magnetic-field induced mechanical distortions of the sample environment \cite{SM}. Data in (A) and (B) were obtained using different instruments at the same beamline \cite{SM}.}
\end{figure}

Another feature of the charge order in YBCO, as well as in \lsco~(where the CO is short range), is that both its correlation length and integrated intensity are suppressed below $T_c$ -- a clear indication of a competition between ordered states \cite{Ghiringhelli_CDW_2012, Chang_CDW_2012, Thampy_2014, Croft_PRB_2014, Christensen_2014}. Figure\,\ref{fig:4}A shows $\theta$-scans measured below and above $T_c$ showing that the CO peak is remarkably insensitive to superconductivity in NCCO. This behavior is not without precedent, since signatures of competition in the temperature dependence of other hole-doped cuprates are not clearly present in the RXS data \cite{daSilvaNeto_CDW_2014, Tabis_2014}. 

The relationship between the charge order and superconductivity can also be probed by measurements in applied magnetic fields. In the case of YBCO, fields up to $18$\,T enhance the CO peak \cite{Chang_CDW_2012, Santi_2014, Huecker_2014}, and at higher fields, above the superconducting upper critical field, $H_{c2}$, the CO becomes long-range and three-dimensional \cite{Wu_2015, Gerber_2015, Chang_2015} -- again indicating a competition between superconductivity and CO. Therefore, given that NCCO has a much lower $H_{c2}$ ($\sim 10$\,T at $0$\,K) \cite{Hidaka_1989, Wang_Science_2003} than hole-doped cuprates, one might expect that an even smaller magnetic field enhanced the CO signal. Measurements in the presence of a magnetic field are more challenging due to field-induced mechanical distortions of the sample environment which can cause significant modifications to the background of the $\theta$-scans \cite{SM}. Nevertheless, Fig.\,\ref{fig:4}B shows that at $10$\,K, below $T_c$, no appreciable difference is seen in the scattering peak up to $6$\,T (the upper limit allowed by our instrument). This finding suggests an insensitivity of the CO to superconducting order, consistent with both the zero-field data across $T_c$ (Fig.\,\ref{fig:4}A) and the doping dependence (Fig.\,\ref{fig:3}B).

\section{Discussion}

Our comprehensive data for the doping, temperature, and magnetic field dependence of the CO in NCCO and LCCO allows us to make a direct comparison to YBCO, a material for which CO has been extensively characterized over the past few years. We find that the CO in NCCO differs from the behavior of its YBCO analog in three ways: (a) it is insensitive to superconductivity, (b) has a small, temperature-independent correlation length, and (c) can be present up to very high temperatures. The YBCO experiments have been interpreted as evidence for competition between different many-electron ground states. Clearly this description does not apply to the CO in NCCO. 

We consider two possible scenarios for the interpretation of our data. First, we refer to recent theoretical work that proposes that disorder-induced Friedel oscillations are responsible for the observation of RXS peaks, akin to quasiparticle interference modulations seen by STM \cite{Abbamonte_2012, Dalla_Torre_NJP}. Though the short, temperature-independent correlation length observed in our measurements might be explained by the length scale of the disorder potential in this scenario, and miscrostructural defects that can act as potential pinning sites are indeed present in NCCO  \cite{Mang_NCCO_defects}, at this point only detailed spatially resolved measurements might be able to validate this scenario. As for the second scenario, we note that in our experiments we do not have the ability to select the energy of the scattered photons, and our measurements should be regarded as energy-integrated. Therefore, as an alternative interpretation, it is possible that the CO peaks in NCCO and LCCO, and possibly even in Bi-based cuprates and Hg1201, are a signature of CO fluctuations rather than static order, in a manner resembling thermal diffuse scattering from soft but weakly temperature dependent lattice vibrations. Indeed, even though the observed softening of the bond stretching phonon mode  \cite{dAstuto_2002} is likely related to the CO, static order is expected to induce a corresponding lattice distortion in NCCO, which has not yet been observed – unlike the case for YBCO where hard x-ray scattering also detects the CO  \cite{Chang_CDW_2012, Huecker_2014}. In this context, it is also worth mentioning that a fluctuating order competing with superconductivity has been observed in NCCO ($x = 0.156$) \cite{Hinton_TRR_2013}. Although a correspondence between this fluctuating order and the CO studied here cannot yet be concluded, we raise the possibility that the competition between superconductivity and the CO can only be observed by separating its inelastic signal from the impurity-pinned quasielastic component. At this point, only more detailed studies of the electronic excitation spectrum will be able to resolve the energy structure of the CO in electron-doped cuprates.

Clearly, the tendency for the charge degrees of freedom to self-organize is ubiquitous to hole- and electron-doped cuprates, strongly suggesting a common physical origin for these correlations. However, the realization of these charge order correlations into a thermodynamic order that competes for the ground state of the system in zero-field is likely material-dependent. Indeed, the suppression of $T_c$ that occurs in YBCO and LBCO near $1/8$ doping is not reported in other cuprates. Factors that could influence the ground state selection include materials-specific lattice distortions, details of the Fermi surface, and disorder. Finally, our experiments also show that the charge order is not directly linked to either antiferromagnetic correlations or the pseudogap in electron-doped cuprates. Overall, our findings should constrain any future endeavors aiming to provide a microscopic theory of charge order formation in the cuprates.

\section{Materials and Methods}

\subsection{Crystal growth and characterization}

NCCO crystals used for the temperature and doping dependent measurements displayed in Figs.\,\ref{fig:2} and \ref{fig:3} were grown by the Minnesota group in an oxygen flow at a pressure of $4$\,atm using the traveling solvent floating zone technique. For all doping levels (except $x=0.042$) the crystals went through an oxygen-reduction process. The reduction procedure was: $10$ hours in Ar gas at $970^\circ$C, followed by $20$ hours in O$_2$ at $500^\circ$C. The actual doping levels were determined either via inductively coupled plasma (ICP) or by wavelength dispersive spectroscopy (WDS) measurements. Most samples were cut from larger crystals measured in a previous INS measurement \cite{Motoyama_Spin_2007}. NCCO crystals used for the measurements displayed in Fig.\,\ref{fig:4} were grown by the Maryland group using the flux method and annealed for two days at the appropriate temperature to render them superconducting \cite{armitage_progress_2010}. The Ce concentration of the crystals was determined using WDS analysis. The \mbox{$c$-axis} oriented LCCO ($x=0.08$) films were deposited directly on (100) $\textrm{Sr}\textrm{Ti}\textrm{O}_{3}$ substrates by a pulsed laser deposition (PLD) technique utilizing a KrF excimer laser as the exciting light source. The films were typically $100$-$150$\,nm in thickness. The samples were optimized by annealing to give a maximum $T_c$ for the $0.08$ Ce doping and a narrow transition temperature width. The $T_c$ was typically $20 \pm 1$\,K. As typical, the normal state resistivity (in a field above $H_{c2}$) shows a low temperature upturn. These LCCO films are similar to those prepared for other experiments by the Maryland group \cite{Saadaoui_2015}. Further details can be found in the supplementary materials \cite{SM}.

\subsection{Resonant x-ray scattering}
Zero-field RXS measurements were performed at the REIXS beamline of the Canadian Light Source in the $20$ to $380$\,K range and in the UHV diffractometer at the UE46-PGM1 beam line of the Helmholtz-Zentrum Berlin at BESSY II, which allowed measurements down to $10$\,K. Magnetic field measurements were performed at the high-field diffractometer of the same beamline in BESSY II. To maximize the CO diffraction signal all measurements were performed in $\sigma$-geometry (photon polarization in the a-b plane) and with the incoming photon energy tuned to the Cu-$L_3$ edge ($\sim$\,$932$\,eV). The $\theta$-scans were performed with the detector angle fixed at $170^\circ$, resulting in $L$ values near $1.6$\,rlu at the peak positions. NCCO samples grown by the traveling floating zone technique had to be polished for the CO peak to be more clearly observed. Measurements on the polished crystals were consistent with identical measurements on the crystals grown by the flux method, with the latter naturally yielding shiny homogeneous surfaces that were appropriate for our RXS experiments.  
\\

\noindent \textbf{Acknowledgments}: We thank HZB for the allocation of synchrotron radiation beamtime. We thank Yeping Jiang for his assistance to sample characterization and preparation. \textbf{Funding}: Supported by the Canadian Institute for Advanced Research (CIFAR) Global Academy (E.H.d.S.N.); the Max Planck-University of British Columbia Centre for Quantum Materials (E.H.d.S.N. and A.D.); the Killam, Alfred P. Sloan, and NSERC\rq{s} Steacie Memorial Fellowships (A.D.); the Alexander von Humboldt Fellowship (A.D. and M.M.); the Canada Research Chairs Program (A.D. and G.A.S.); and the Natural Sciences and Engineering Research Council of Canada (NSERC), Canada Foundation for Innovation (CFI), and CIFAR Quantum Materials. The work at the University of Minnesota was supported partially by NSF Award 1006617 and by the NSF through the University of Minnesota MRSEC under Award Number DMR-1420013. Work at the University of Maryland was supported by NSF grant DMR-1410665. The beamline REIXS of the Canadian Light Source is funded by CFI, NSERC, National Research Council Canada, Canadian Institutes of Health Research, the Government of Saskatchewan, Western Economic Diversification Canada, and the University of Saskatchewan. \textbf{Author contributions}: E.H.d.S.N, B.Y., M.M., R.S., E.S., F.B., M.Z., and M.B. performed the resonant x-ray scattering measurements. B.Y., J.H., Y.J. Y.L., G.Y., R.L.G., and M.G. synthesized and characterized the materials. E.H.d.S.N. performed the analysis of the RXS data. E.H.d.S.N., B.K., and A.D. wrote the manuscript. All authors read and commented on the manuscript. A.D. was responsible for overall project planning and direction. \textbf{Competing interests}: The authors declare that they have no competing interests.

\bibliographystyle{apsrev4-1}
\bibliography{bib_lib}
\end{document}